\documentclass[aps,nofootinbib,preprintnumbers,citeautoscript]{revtex4}
\usepackage{amsmath,amssymb,epsfig}
\usepackage{enumerate}
\def\be{\begin{equation}}
\def\ee{\end{equation}}

\def\be{\begin{equation}}
\def\ee{\end{equation}}

\def\beq{\begin{eqnarray}}\def\eeq{\end{eqnarray}}

\usepackage{graphicx}

\begin{document}
\title{On the superposition principle in interference experiments }

\author{Aninda Sinha*$^{1}$, Aravind H.~Vijay$^{1,2}$  and Urbasi Sinha*$^{2,3}$\\
\it $^1$ Centre for High Energy Physics, Indian Institute of Science,  Bangalore, India. \\
\it $^2$ Raman Research Institute, Sadashivanagar, Bangalore, India.\\
\it $^3$ Institute for Quantum Computing, 200 University Avenue West, Waterloo, Ontario, Canada.\\
\it{$^\ast$To whom correspondence should be addressed; E-mail:  asinha@cts.iisc.ernet.in, usinha@rri.res.in.}}

\begin{abstract}

The superposition principle is usually incorrectly applied in interference experiments.
 This has recently been investigated through numerics based on Finite Difference Time Domain (FDTD) methods as well as the Feynman path integral formalism. In the current work, we have derived an analytic formula for the Sorkin parameter which can be used to determine the deviation from the application of the principle. We have found excellent agreement between the analytic distribution and those that have been earlier estimated by numerical integration  as well as resource intensive FDTD simulations. 
 The analytic handle would be useful for comparing theory with future experiments. It is applicable both to physics based on classical wave equations as well as the non-relativistic Schr\"odinger equation.

\end{abstract}
\maketitle

%\newpage

%\tableofcontents

\section*{Introduction}
It is not widely appreciated that the superposition principle is incorrectly applied in most textbook expositions of  interference experiments both in optics and quantum mechanics \cite{born,book1, book2, book3}. For example, in a double slit experiment, the amplitude at the screen is usually obtained by adding the amplitudes corresponding to the slits open one at a time. However, the conditions described here correspond to different boundary conditions (or different Hamiltonians) and as such the superposition principle should not be directly applicable in this case. This incorrect application was pointed out in a physically inaccessible domain by \cite{yabuki} and in a classical simulation of Maxwell equations by \cite{draedt}. More recently, \cite{sinhas} dealt with the quantification of this correction in the quantum mechanical domain where the Feynman path integral formalism  \cite{feynman} was used to solve the problem of scattering due to the presence of slits . According to the path integral formalism, the probability amplitude to travel from point A to B should take into account all possible paths with proper weightage given to the different paths. In the nomenclature used, paths which extremize the classical action are called ``classical" paths whereas paths which do not extremize the action are called``non-classical" paths. In the usual Fresnel theory of diffraction, the assumption is that the wave amplitude at a particular slit would be the same as it would be away from the slits. Adding to Fresnel theory, we take into account a higher order effect and also account for influence by waves arriving through neighboring slits. This way the naive application of the superposition principle is violated. This discussion makes it clear that our approach is equally applicable to physics described by Maxwell theory and Schr\"odinger equation--see supplementary material of \cite{sinhas} for further details.

In \cite{sorkin, sinhas, usinha}, the normalized version of the Sorkin parameter $\kappa$ (defined later) was estimated. This would be zero if only the classical paths contribute and would be non-zero when the non-classical paths are taken into account. The proposed experiment in \cite{sinhas} to detect the presence of the non-classical paths uses a triple slit configuration as shown in fig.(1). However, \cite{sinhas} was restricted to semi-analytic and numerical methods. The analysis using path integrals was restricted to the far field regime {\it i.e.,} the Fraunhofer regime in optics and considered cases in which the thickness of the slits is negligible. Only the first order correction term was considered in which paths of the kind shown in the inset of fig.(1) contribute. In the current work, we have derived an analytic formula for $\kappa$ as a function of detector position in the Fraunhofer regime. We find that the quantity $\kappa$ is very sensitive to certain length parameters. Thus having an analytic handle is very important as this makes it a much more accessible quantity to experimentalists. This would enable experimentalists to have a feel for how errors in the precise knowledge of various parameters can affect the $\kappa$ distribution on the detector plane thus making it easier to compare theory with experiments. The analytic formula now makes the understanding of the deviation from the naive application of the superposition principle more tractable, shedding more light on the``black-box" like understanding that numerical simulations could afford. 
We have showed that the analytic formula gives us an excellent match with both photon and electron parameters used in \cite{sinhas}. Moreover, it compares very well with \cite{draedt} where a classical simulation of Maxwell equations using Finite Difference Time Domain (FDTD) methods was done. An important point to note here is that an FDTD simulation of $\kappa$ in \cite{draedt} needed several %900GB of memory and about 12 hours of computation time using 8192 processors of the IBM BlueGene/P at the J¨ulich Supercomputing Centre resulting in 
days of computation time of a supercomputer and several terabytes of memory, while our analytic formula gives us a $\kappa$ distribution almost immediately on a standard laptop using {\it Mathematica}. Of course, an FDTD simulation will be able to capture effects due to material properties of the absorber as well as be applicable to near field regimes. However, our analytic approximation now makes it easy to describe the effect of non-classical paths in different experimental scenarios without having to go through resource intensive detailed numerics.

 In addition to the analytic handle on $\kappa$, we have done a path integral based simulation using a different numerical approach from \cite{sinhas} based on Riemannian integration. This has enabled us to include both far field and near field regimes in our analysis. We have also verified the effect of increasing the number of kinks in the non-classical paths. %({\bf you mention one line on this in the analytical formula part but I think it is important to discuss this in a few more lines somewhere, either there or in the helmholtz equation section which may be more appropriate: Leave this to you}). %
Our current results now make the experimental conditions required less restrictive in terms of  length parameters other than of course providing further verification for the results obtained in \cite{draedt} and \cite{sinhas} .

\begin{figure}
\centering
\includegraphics[scale=0.400]{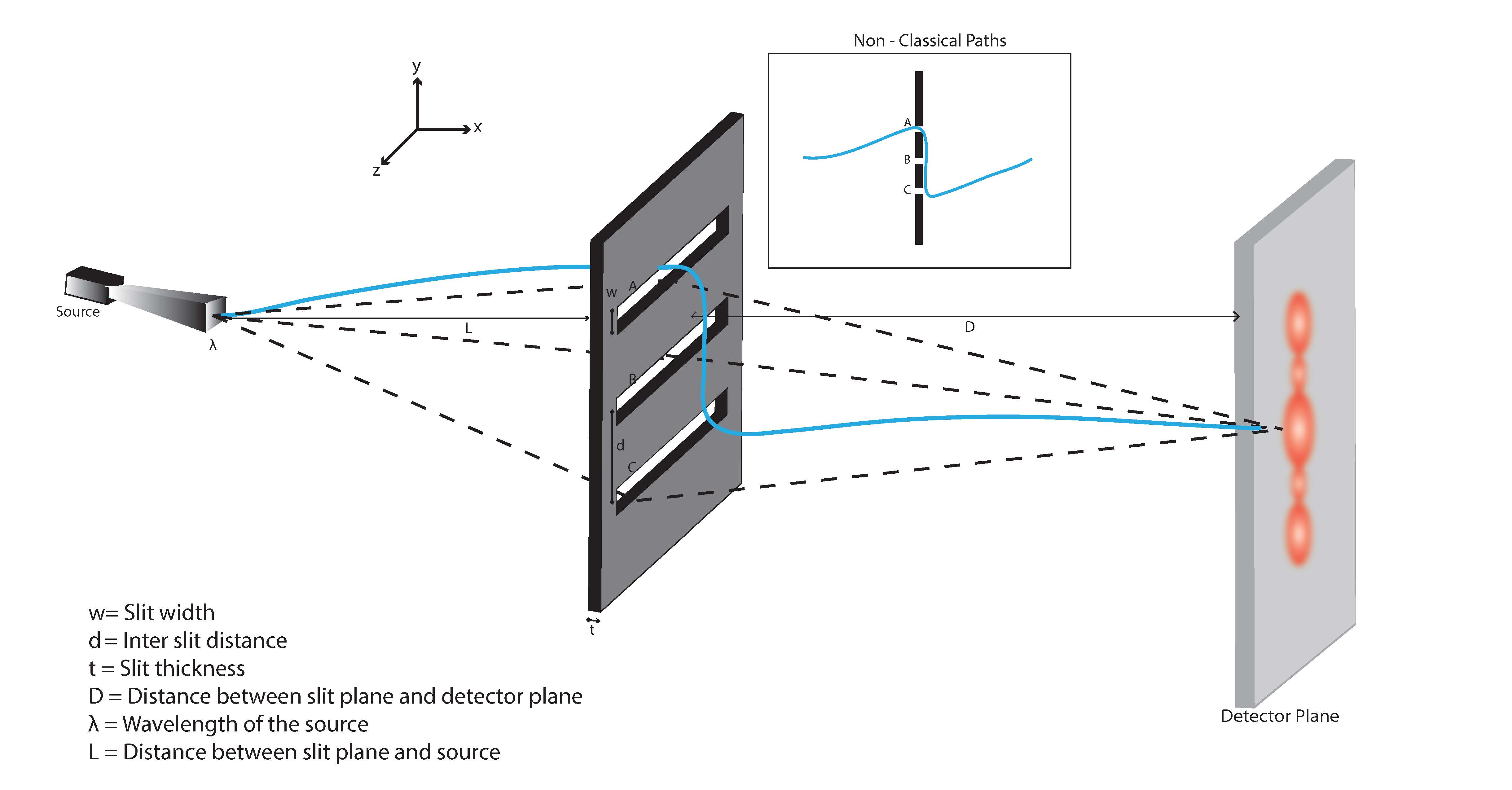}

\caption{The triple slit set-up with a representative non-classical path. A grazing path has been illustrated in the inset where a path enters slit A, goes to slit C, just enters it and then goes to the detector. We integrate over the widths of slits A and C for this path.}
\label{}
\end{figure}

%\begin{figure}
%\centering
%\includegraphics[scale=0.500]{image1.pdf}
%\caption{The two slit experiment. The common assumption is that the wave function corresponding to both slits open {\it i.e.,} $\psi_{AB} = $ sum of wave functions corresponding to the slits open one at a time {\it i.e.,} $\psi_A + \psi_B$}
%\label{}
%\end{figure}
  %
\section*
{The Sorkin parameter $\kappa$}

Consider the triple slit configuration shown in fig.(1). Let the three slits be labelled A, B and C respectively. The wave function corresponding to slit A being open is $\psi_A$, that corresponding to slit B being open is $\psi_B$ and that corresponding to slit C being open is $\psi_C$. Similarly, for both A and B open, it will be $\psi_{AB}$, for A, B and C open, it will be $\psi_{ABC}$ and so on. Now, a naive application of the superposition principle will dictate that $ \psi_{AB} = \psi_A +\psi_B$ . However as pointed out in references \cite{yabuki,draedt,sinhas}, this approximation is strictly not true as the situations described correspond to three different boundary conditions and the superposition principle cannot be applied to add solutions to different boundary conditions to arrive at a solution for yet another one. This leads to a modification of the wave function at the screen which now becomes:
\begin{eqnarray}
     \psi_{AB} = \psi_A + \psi_B + \psi_{nc}\,
\label{eqn:superposition}
   \end{eqnarray}
where $\psi_{nc}$ is the contribution due to the kinked {\it {\it i.e.,},} non-classical paths.  \cite{sinhas} uses the Feynman path integral formalism to quantify the effect due to non classical paths in interference experiments which helps in getting an idea about the correction $ \psi_{nc}$ . The normalized version of the Sorkin parameter called $\kappa$ was used to propose experiments which can be done to measure such deviations. The quantity $\kappa$ has a special symmetry in its formulation which ensures that it evaluates to zero in the absence of any contribution from the correction term but assumes a finite non-zero value when the correction term is present. The numerator of $\kappa$ which we call $\epsilon$ is defined as follows:
\begin{eqnarray}
\epsilon = I_{ABC}-(I_{AB}+I_{BC}+I_{CA})+(I_A+I_B+I_C) \,
\label{eqn:sorkinnumber}
\end{eqnarray}
where $I_{ABC}$ is the probability or the intensity at the screen when all three slits are open, $I_{AB}$ is the intensity at the screen when slits A and B are open and so on. Taking into account eqn. (\ref{eqn:sorkinnumber}), $\kappa$ is defined as follows:
\begin{eqnarray}
\kappa = \frac{\epsilon}{\delta} \,
\label{eqn:kappa}
\end{eqnarray}

where $\delta$ is defined as the value of the intensity at the central maximum of the triple slit interference pattern.
If the correction term in eqn.(\ref{eqn:superposition}) is not taken into account, then $\kappa$ will evaluate to zero from algebra (This is under the assumption that Born's rule for probability i.e. Probability $\propto |\psi|^2$ is true). The presence of the correction term $\psi_{nc}$ makes $\kappa$ manifestly non-zero (as explicitly shown in the next section) thus making it a perfect tool to investigate such correction effects to the application of the superposition principle in interference experiments. One has to note here that such correction effects are not a purview of quantum mechanics alone. Even if one considers classical Maxwell equations and then applies the different boundary conditions corresponding to slits being open one at a time and then all together, one is able to get a difference in the two situations as per eqn.(\ref{eqn:superposition}). This was shown through FDTD solutions in \cite{draedt}. In \cite{sinhas}, we used Feynman Path integral formalism to analyze the problem and this made the analysis applicable to the domain of single particles like electrons, photons and neutrons. One has to note that the path integral analysis is also applicable to the classical domain as we had used the time independent Helmholtz equation propagator which is applicable to both Maxwell equations and for instance the time independent Schrodinger equation.

\section*{Analytic approximation for $\kappa$ in the thin slit case}
In this section we will discuss how to get an analytic expression for the normalized Sorkin parameter $\kappa$ in the thin slit approximation ($t$ in figure 1 is much smaller than other length parameters in the problem) and in the Fraunhofer limit. We begin by reviewing the logic in \cite{sinhas}.
The wave function at the screen gets contributions from several paths. We can subdivide the paths into ones which involve the classical straight paths going from source to slit and then slit to screen, ones which go from source to slit P then from slit P to slit Q and then to the screen and so on. The second category of paths can further have kinks in them. We will assume that the dominant path from source to slit P is the straight line one and then from slit P to slit Q is also a straight line one (since the classical path from P to Q is a straight line one) and from slit Q to the screen is again the straight path. This approximation seems reasonable to us and we have confirmed this by explicitly adding kinks to these paths and numerically checking that their contribution is negligible.
Keeping this discussion in mind we can for example write the wave function at the screen as:
\begin{eqnarray}
\psi &=& \psi_A+ \psi_B+\psi_C+\psi_{A,B}+\psi_{A,C}+\psi_{B,A}+\psi_{B,C}+\psi_{C,A}+\psi_{C,B} \,, ~~~~A,B,C {\rm~open}\\
\psi &=& \psi_A+ \psi_{A,B}+\psi_{B,A}\,,~~~~~~~~~~~~~~~~~~~~~~~~~~~~~~~~~~~~~~~
~~~~~~~~~~~~~~~~~~~~A,B {\rm~open}\\
\psi &=& \psi_A\,,~~~~~~~~~~~~~~~~~~~~~~~~~~~~~~~~~~~~~
~~~~~~~~~~~~~~~~~~~~~~~~~~~~~~~~~~~~~~~~~~A {~\rm open}
\end{eqnarray}
where $\psi_A$ denotes the classical contribution, $\psi_{A,B}$ denotes the non-classical contribution corresponding to a path going from source to A, A to B and then B to detector, $\psi_{A,C}$ denotes the non-classical contribution corresponding to a path going from source to A, A to C and then C to detector. Similarly for the other cases.
Using this it is easy to check that the numerator in $\kappa$ becomes
\begin{equation}
\epsilon \approx 2 {\rm Re}[\psi^*_A(\psi_{B,C}+\psi_{C,B})+\psi^*_B(\psi_{A,C}+\psi_{C,A})+
\psi^*_C(\psi_{A,B}+\psi_{B,A})]\,,
\end{equation}
where we have ignored the second order terms like $\psi_{B,C}\psi^*_{C,A}$ since these turn out to be much smaller compared to the terms displayed above.  An important point to note here is that we could replace $\psi_P$ by $\psi_P+\psi^{k}_P$ where $\psi^{k}_P$ denote contributions due to kinks in the paths going from the source to the slit P and then to the detector. However these contributions cancel out in $\kappa$ as can be easily checked. This is one of the main reasons why $\kappa$ the triple-slit set up is preferred over the double-slit interference $I_{AB}-I_A-I_B$ in our discussion of non-classical paths, since in the latter, contributions from $\psi^{k}_P$ cannot be ignored and these are difficult to estimate. 
As in \cite{sinhas} we will use the free particle propagator for a particle with wave number $k$ going from $\vec{r}$ to $\vec{r'}$
\begin{equation}
K(\vec{r},\vec{r'})=\frac{k}{2\pi i} \frac{e^{ik|\vec{r}-\vec{r'}|}}{|\vec{r}-\vec{r'}|}\,.
\end{equation}
Here the normalization factor has been fixed by demanding the composition rule following \cite{landau} $K(\vec{r_1},\vec{r_3})=\int d {\vec r_2^\perp} K(\vec{r_1},\vec{r_2})K(\vec{r_2},\vec{r_3})$ where the integration is over a plane perpendicular to $\vec{r_1}-\vec{r_3}$. We will consider the evolution of the wavefunction from the source to the detector which is given in the Feynman path integral formulation of quantum mechanics by summing over all paths that go from the source to the detector. Any path can be thought to be made by integrating small straight line propagators.
The $y$-coordinate extents of the slits are $A: d-w/2, d+w/2$, $B:-w/2,w/2$ and $C:-d-w/2,-d+w/2$. Namely the centre  to centre distance is $d$ and the width is $w$. We will assume that the slit has negligible thickness and hence there is no need for an $x$-integration. Further as argued in \cite{sinhas}, after the stationary phase approximation, the $z$-integration along the height of the slit in the numerator and denominator are the same and hence we will not have to worry about this either and we will drop the $z$ integrals from the beginning. We will further assume a Fraunhofer regime so that the source to slit distance $L$ and slit to detector/screen distance $D$ are both large compared to any other scale in the problem.
Using the results of \cite{sinhas}, we have
\begin{eqnarray}
\psi_B &=&-\gamma \frac{k}{4\pi^2} \int_{-\frac{w}{2}}^{\frac{w}{2}}dy e^{ik[\frac{y^2}{2L}+\frac{(y_D-y)^2}{2D}]} \approx -\gamma \frac{k}{4\pi^2} \int_{-\frac{w}{2}}^{\frac{w}{2}}dy e^{-ik\frac{(y_D y)}{D}} \,,\\
\psi_A &=& e^{-i k d\frac{y_D}{D}} \psi_B \,,\\
\psi_C &=& e^{i k d\frac{y_D}{D}} \psi_B \,,
\end{eqnarray}
where $\gamma=e^{ik(L+D)}/LD$. In the first line, we have dropped the quadratic terms since $y^2/L, y^2/D$ are very small in the domain of integration while we have retained the linear term $y y_D/D$ in order to be able to compute what happens at the detector screen far away from the centre.
After doing the stationary phase approximation as explained in \cite{sinhas}, we find that 
\begin{equation}
\psi_{P,Q}\approx \gamma i^{3/2} (\frac{k}{2\pi})^{5/2}\int dy_1 dy_2 |y_2-y_1|^{-1/2} e^{i k |y_2-y_1|-i k y_2 \frac{y_D}{D}}\,,  
\end{equation}
where the $y_1$ integral runs over slit P and the $y_2$ integral runs over slit Q. At this stage we observe that if $|y_2-y_1|$ is large in the domain of integration, then we can approximate the $y_1,y_2$ integrals by retaining the leading order term obtained by integrating by parts--this is the standard technique used to approximate such integrals leading to an asymptotic series. Explicitly, we use denoting $y_D/D=\theta$ and rescaling $y_1, y_2$ by $k$ we have
\begin{eqnarray} 
\int_a^b dy_1 \int_p^q dy_2 && \!\!\!\!\!\!\!\!\!\!\!|y_2-y_1|^{-1/2} e^{i |y_2-y_1|-i y_2 \theta}
= \Theta(b-y_2) \frac{e^{i(b-y_2)-i y_2 \theta}}{(b-y_2)^{1/2}} {\bigg |}_p^q+ \Theta(y_2-b) \frac{e^{-i(b-y_2)-i y_2 \theta}}{(y_2-b)^{1/2}}{\bigg |}_p^q \nonumber \\
&-& \Theta(a-y_2) \frac{e^{i(a-y_2)-i y_2 \theta}}{(a-y_2)^{1/2}}{\bigg |}_p^q - \Theta(y_2-a) \frac{e^{-i(a-y_2)-i y_2 \theta}}{(y_2-a)^{1/2}}{\bigg |}_p^q+O\left(\frac{1}{|b-y_2|^{3/2}},\frac{1}{|a-y_2|^{3/2}}\right)\,, \label{asym}
\end{eqnarray} 
where $\Theta(x)$ is the Heaviside step function and takes into account the modulus sign in the integrand (we have assumed that $\theta \ll 1$).
Using this we finally find (here all length variables have been rescaled by $k$)
\begin{equation}
\kappa(\theta)=\frac{1}{9\sqrt{2\pi d}}\frac{\sin w\theta}{w^2 \theta} f(d,w,\theta)\,,
\end{equation}
where
\begin{eqnarray}
f(d,w,\theta)&=&\left[2\cos(2d\theta)\cos(d-\pi/4)+\sqrt{2}\cos(d\theta)\cos(2d-\pi/4)+2\cos(d\theta)\cos(d-\pi/4)\right] \nonumber\\ &-&2\sqrt{\frac{d}{d-w}}\frac{\cos((d-w)\theta/2)}{\cos(w\theta/2)}\cos(3d\theta/2)\cos(d-w-\pi/4) \nonumber \\&-& 2\sqrt{\frac{d}{d+w}}\frac{\cos((d+w)\theta/2)}{\cos(w\theta/2)}\cos(3d\theta/2)\cos(d+w-\pi/4)\nonumber \\&-&\sqrt{\frac{d}{2d-w}}\frac{\cos((2d-w)\theta/2)}{\cos(w\theta/2)}\cos(2d-w-\pi/4)\nonumber \\&-&\sqrt{\frac{d}{2d+w}}\frac{\cos((2d+w)\theta/2)}{\cos(w\theta/2)}\cos(2d+w-\pi/4)\,\nonumber\\
\end{eqnarray}
In deriving the above expression, we have put in an extra factor of $1/4$ that arises from inclination factors as argued in \cite{sinhas}.
One important cross check that this formula satisfies is that $\kappa (\theta)$ becomes zero when $\it{w}$ goes to zero. We point out that the above result is a very good approximation to $\kappa$ in the limit when $d\gg w \gg 1$ since the term that is neglected in eqn.(\ref{asym}) is $O(\lambda^{3/2}/(2\pi)^{3/2}(d-w)^{3/2})$ after reinstating factors of $k$.

Using this approximate expression we can compare with the results of numerical integration in \cite{sinhas}. This is shown in fig.2(a) and fig.2(b). We find that the agreement with the numerics is excellent in the far field regime.
% This is to be expected since in deriving the approximate analytic expression we assumed that $\theta\ll 1$ in radians. To get a better agreement we can retain the next term in the asymptotic expansion but we will not need this in what follows.%

\begin{figure}[ht]
\includegraphics[scale=0.700]{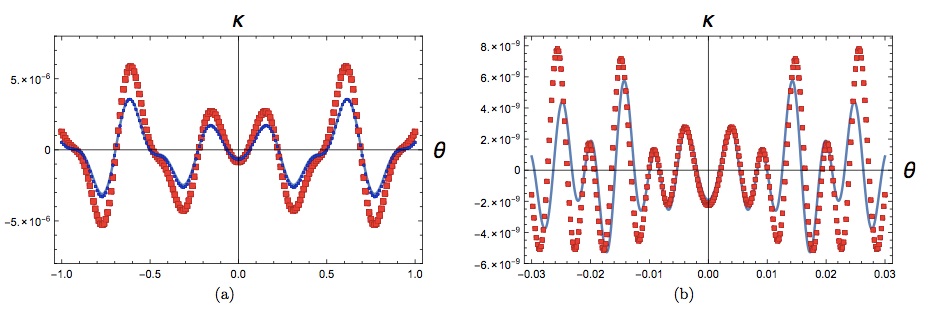}
%\begin{figure}[ht]
%\begin{tabular}{c c}
%\includegraphics[scale=0.700]{3a.jpg}
%& 
%~~~~\includegraphics[scale=0.700]{3b.jpg}\\
%(a) & (b)
%\end{tabular}
\caption{Comparison between numerics in \cite{sinhas} and the analytic approximation. Figure on the left shows $\kappa$ as a function of angle $\theta$ (where $\theta = \frac{y_D}{D} \frac{180}{\pi}$ in degrees) for the photon parameters {\it i.e.,} slit width = 30$\mu m$, inter- slit distance =  100$\mu m$ and wavelength of incident photon = 810nm. The red dotted line indicates the result of numerical integration where source-slit distance and slit-detector distance = 18.1cm as in \cite{sinhas}. This corresponds to a Fresnel number of 0.006. The blue line indicates the result of application of the analytic formula in eqn.(14) and the blue dots show the result of numerical integration when the Fresnel number has been adjusted to 0.0002. This implies that in the far field regime, the analytic formula and numerical integration results show perfect overlap. We find that for Fresnel number $\lesssim 0.001$ leads to a discrepancy of $\lesssim 10\%$ at the centre. Figure on the right shows $\kappa$ as a function of detector position for the electron parameters \cite{electron} {\it i.e.,} slit width = 62nm, inter-slit distance = 272nm, distance between source and slits = 30.5cm and distance between slits and detector =24cm and de Broglie wavelength of incident electrons = 50pm.The red dots indicate the result of numerical integration as per \cite{sinhas}. The blue line indicates the result of application of the analytic formula in eqn.(14). The Fresnel number is 0.0002. }
\end{figure}

Now using the analytic expression, we can derive a bound on $\kappa$ in the regime $k w\gg 1, d\gg w$. By setting the trigonometric functions to their maximum value and adjusting all the relative signs to be the same, we find after reinstating factors of $k$
\begin{equation}
|\kappa_{max}| \approx 0.03 \frac{\lambda^{3/2}}{d^{1/2} w}\,.
\end{equation}
In all examples we have numerically verified that this is a strict upper bound. It will be a useful simple formula to remember.
%%%%%%%%%%%%%%%%%%%%%%%%%%%%%
\section*{Comparison with FDTD}
We can use the analytic expression to compare with the FDTD results in \cite{draedt}. Although the FDTD simulations were done for non-zero thickness for the slits and for non-ideal materials, we will find that the analytic formula agrees remarkably well with the FDTD results. The reason for this agreement is the following. One can repeat the steps outlined above but now with thickness. To handle the thick slit case, we can consider two slit planes instead of one where the separation between the two planes is given by the thickness of the slit. Then there are paths that reach from the source to the first slit plane, from the first plane to the second plane, and finally from the second plane to the detector. We can as before use the stationary phase approximation. In the end we find that again the $z$-integrals cancel out and we are left with expressions of the form
\be
K^A_{nc}=\gamma (\frac{k}{2\pi})^3 e^{i\pi/2} \int dy_1 dy_2 dy_3 \frac{\exp[i k[\frac{y_1^2}{2L}+\sqrt{(y_2-y_1)^2+t^2}+\frac{(y_D-y_3)^2}{2D}+|y_3-y_2|]]}{|y_3-y_2|^{1/2}[(y_2-y_1)^2+t^2]^{1/4}}\,,
\ee
where $t$ here is the thickness. $y_1,y_2$ are the y-coordinates on the first and second slit plane respectively involving slit A and $y_3$ is on the second slit plane involving slits B or C--this denotes a path where the kink in the path occurs at the second slit plane.% {\bf this line is not very clear: see if the above reads better}.%
 Now there are two observations to make. First, there should also be a contribution from a path that has a kink in the first slit plane. When the thickness is small ($t\sim \lambda$) then these two paths will approximately be in phase and hence there will be an overall factor of $2^2=4$ in $\kappa$ compared to the thin slit approximation. Second, when $t$ is small, the factor $1/[(y_2-y_1)^2+t^2]^{1/4}$ will be sharply peaked around $y_2=y_1$ and hence the result of the $y_2$ integral will lead to an expression which is the same as in the thin slit case. 
Now if we wanted to compare with the FDTD simulations in \cite{draedt}, we note that the material making the slits in the simulations was considered to be steel with a complex refractive index. The effect of the imaginary part of the refractive index is to make the effective slit width bigger compared to the idealized scenario we are considering. By considering a slightly bigger $w$, we find that the agreement of the analytic expression for $\kappa$ with the FDTD simulation for $d=3\lambda,w=\lambda, t=4\lambda$ as considered in \cite{draedt} is remarkably good as shown in fig.(3). The complex refractive index for steel as used in \cite{draedt} for FDTD simulation is $n=n_R+n_I i=2.29+2.61 i$. Using the fact that the wave gets attenuated by $\exp(-2\pi n_I x/\lambda)$ \cite{griffiths} at a distance of $x$ inside the material, we find that for $x=0.075 \lambda$ the attenuation factor is $30\%$. This gives an effective increase in the slit width which we will take to be $2\times0.075\lambda=0.15\lambda$.
\begin{figure}[ht]
\includegraphics[scale=0.400]{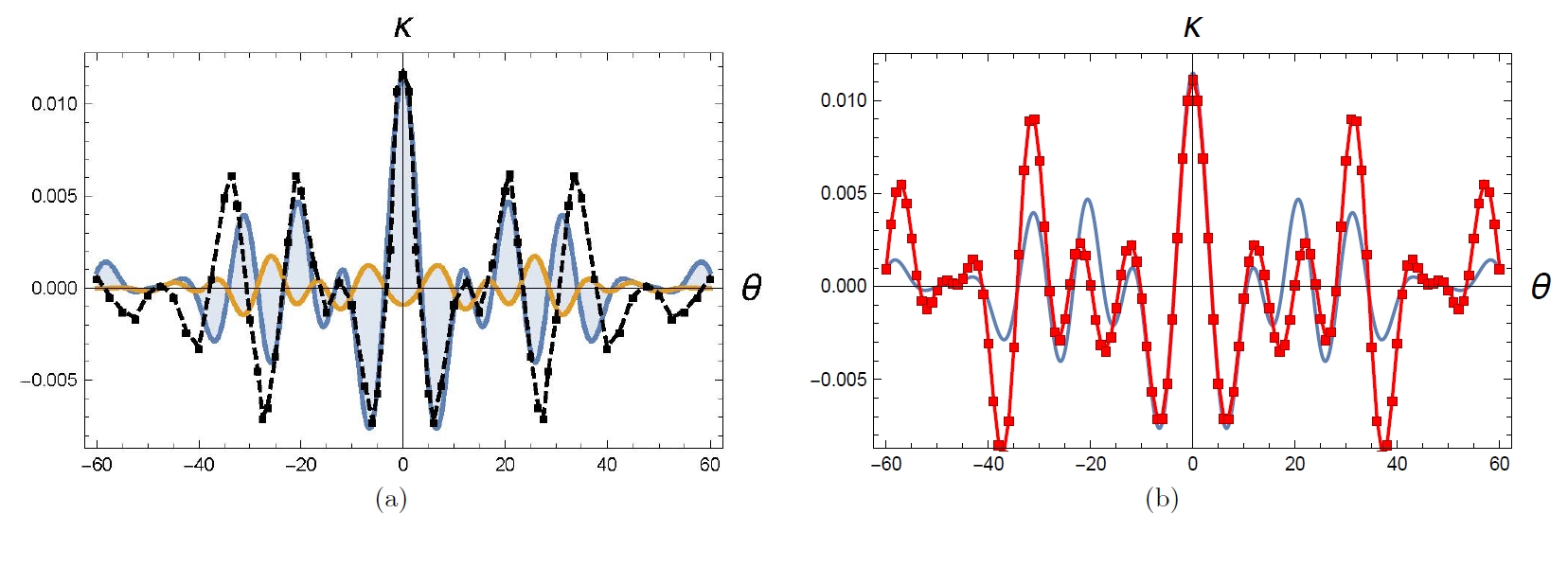}
%\begin{tabular}{c c}
%\includegraphics[scale=0.700]{de1.jpg}
%& 
%~~~~\includegraphics[scale=0.700]{decomp.pdf}\\
%(a) & (b)
%\end{tabular}
\caption{Comparison with the FDTD simulations in \cite{draedt} for the $d=3\lambda,w=\lambda, t=4\lambda$ case. In the figure on the left, the black dots indicate the FDTD values which have been read off from fig.(2b) in \cite{draedt}. The orange line indicates the analytic expression while the blue line which leads to an agreement with the FDTD result is the analytic expression with $d=3\lambda,w=1.15\lambda$. This choice of $w$ in the analytic expression has been justified in the text. In the figure on the right, we compare the analytic formula with the numerical integration as in \cite{sinhas} for the $d=3\lambda,w= 1.15\lambda, t=4\lambda$ case. The red dots indicate the result from numerical integration while the blue line indicates the analytic expression. The good agreement, especially close to the central region justifies our usage of the analytic approximation for this choice of parameters.}
\label{}
\end{figure}

%\begin{figure}
%\centering
%\includegraphics[scale=0.700]{de1.jpg}
%\caption{Comparison with the FDTD simulations in \cite{draedt} for the $d=3\lambda,w=\lambda, t=4\lambda$ case. The black dots indicate the FDTD values which have been read off from fig.(2b) in \cite{draedt}. The orange line indicates the analytic expression while the blue line which leads to an agreement with the FDTD result is the analytic expression with $d=3\lambda,w=1.15\lambda$. This choice of $w$ in the analytic expression has been justified in the text. }
%\label{}
%\end{figure}

\section*{ The Sorkin parameter in the Fresnel regime}

In reference \cite{sinhas} as well as in deriving the analytic approximation in our current work, we needed to be in the Fraunhofer regime. This enabled us to expand the propagator distance for example in eqn.(9) which was crucial in the simplifications arising from the stationary phase approximation namely the integral over the height of the slits cancelled between the numerator and denominator in $\kappa$. However, in order to consider the Fresnel regime, we can no longer appeal to this simplification and will need to consider a different numerical approach. While FDTD can enable us to address the same question, as pointed out in the Introduction, it is computationally resource intensive. The approach we will outline below is more efficient in addressing this issue.
We will use the common technique for numerical integration which is Riemannian integration \cite{C}. The technique involves dividing a certain domain into many smaller sub-domains and assuming that the integrand function is constant across the domain. One then sums up the constants multiplied by the area of the sub-domains to get the integral of the function over the whole domain. Our code to evaluate $\kappa$ was written in the C++ programing language. We retained the exact propagator distances and integrated over the length of the slit along the z-axis in figure 1. We used the same parameters that were used to generate fig.(3a) of reference \cite{sinhas} and in addition chose the height to be 300$\mu$m. 

%\subsection*{Far field and near field regimes}
\begin{figure}
\centering
\includegraphics[scale=0.700]{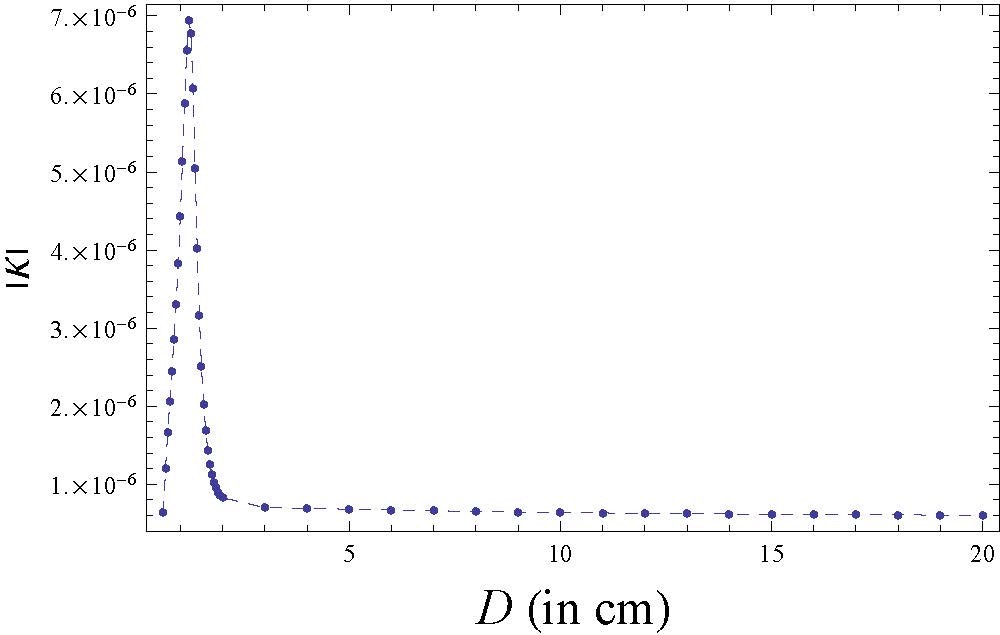}
%\caption{}
\caption{$|\kappa|$ (at the central maximum of the triple slit interference pattern) as a function of distance between slit plane and detector screen plane $D$. The parameters used are  slit width $w = 30\mu$m, inter slit distance $d= 100\mu$m, height of the slit = $300\mu$m, incident wavelength $\lambda$ = 810nm, source to slit plane distance $L= 20cm$. }
\end{figure} 
Figure 4 shows $|\kappa|$ as a function of distance between slit plane and detector plane $D$. We find that using this approach, the value of $|\kappa|$ at $D = 20cm$ is $|\kappa|\approx 6 \times10^{-7}$ while the analytic formula gives $\approx 5.6\times10^{-7}$ which means a deviation of around 7$\%$. Thus, already for a Fresnel number of 0.005 (which corresponds to D = 20cm), the agreement between our numerical integration approach and analytic approximation is very good. As the distance between the slit plane and the detector plane decreases,  the value of $|\kappa|$ starts increasing which can be explained by the decrease of the value of the denominator of $|\kappa|$. However, the sudden dip at very close distances may be an artefact of our approximation and the fact that the paraxial approximation breaks down in the extreme near field regime. 

Can we use the new numerical approach to compare with FDTD results? In our current numerical approach, we have made certain assumptions which are summarized next. The first assumption is of a steady source which follows scalar electrodynamics. This approximation will break down in case of polarized radiation but in construction of the quantity $\kappa$ for unpolarized light, the polarization sums cancel in the numerator and denominator. Moreover, induction of currents in materials used for building the apparatus and scattering of radiation are not accounted for in the above derivations {\it i.e.,} we have assumed that our material is a perfect absorber. This approximation will break down when the material scatters radiation to a significant effect and behaves as a secondary source due to induction. We have also used an approximate form for Kirchoff's integral theorem \cite{born} whereby it is important that length scales in the problem are much larger than the wavelength of incident radiation. One could use the complete Kirchoff's boundary integral in order to investigate problems where the length scales are comparable to wavelength.  A final approximation used in this section is that of paraxial rays. This one breaks down when the distance between the slit plane and source or screen is not large compared to the vertical position on the screen or the slit width {\it i.e.,} when we want to plot $\kappa$ as a function of detector positions which are very far from the central region. We leave more careful investigation of the Fresnel regime for the comparison between the Riemannian integration based technique outlined above and the resource intensive FDTD approach for future work.

\section*{Discussion}
In this paper, we have derived an analytic expression for the Sorkin parameter $\kappa$ which has been used to quantify deviations from the naive application of the superposition principle in slit-based interference set-ups. Our main formula in eqn.(14) can be trusted in the Fraunhofer regime and in a thin-slit approximation. When the thickness of the slit plane is not too big, we have given an argument on how to use our analytic formula which led to impressive agreements with the FDTD simulations of \cite{draedt}  as well as numerical integrations of \cite{sinhas}. In the future, it will be interesting to develop systematics of the thick-slit scenario following some of the techniques used in this paper. One important point to note is that in the final expression for $\kappa$ \i.e. eqn.(14), $\hbar$ appears only indirectly through the de Broglie wavelength. This is in keeping with our claim that our analytic formula should be applicable for both Maxwell's equations as well as the Schr\"odinger equation. The non-zeroness of $\kappa$ is essentially due to boundary condition considerations and should affect both classical as well as quantum physics. In existing experimental results in literature which measure $\kappa$ for example \cite{usinha,weihs,laflamme,pra}, the experimental inaccuracies have prevented us from concluding that $\kappa$ is non-zero. As is easy to see in our analytic formula, $\kappa$ is very sensitive to experimental parameters. Future experimental attempts will benefit from our analytic handle as it would be much easier to compare experimental data with theoretical expectations.  One has to note here that the quantity $\kappa$ has been measured previously to test for a possible deviation from Born rule. So, what do our findings imply for using $\kappa$ as a test for Born rule? $\kappa$ should be used for a Born rule test in experimental situations where the non-zeroness due to the correction to the superposition principle is very small. For instance, a set-up like in \cite{weihs} has a very small correction from non classical paths and could be a good experiment still to test Born rule. Thus, any potentially detectable violation of the Born rule should be bigger than that due to non-classical paths and any future test should take this into account.

\section*{Acknowledgments} We thank Joseph Samuel, Barry Sanders, Diptiman Sen and Supurna Sinha for useful discussions and Animesh Aaryan for assistance with figure 1. We thank Anthony Leggett,  Barry Sanders and Diptiman Sen for comments on the draft. US wishes to thank Nicolas Copernicus University, Poland for kind hospitality while a part of the work was being carried out. AS acknowledges partial support from a Ramanujan fellowship, Govt. of India.

\section*{Author Contributions}
AS and US formulated the questions, AS performed the analytic calculations with inputs from AHV and US, AHV performed the numerical simulations in the Fresnel regime. AS and US wrote the manuscript and all authors reviewed it.

\section*{Additional information}
The author(s) declare no competing financial interests.

\end{document}